\documentstyle[12pt,epsfig,rotating]{article}
\def\pge{\pagestyle{empty}} \def\pgn{\pagestyle{plain}}
     
\def\spg{\setcounter{page}} 
\def\bd{
\begin{document}} \def\ed{\end{document}}
\def\bmp{\begin{minipage}} \def\emp{\end{minipage}}
\def\bcc{\begin{center}} \def\ecc{\end{center}}     \def\npg{\newpage}
\def\beq{\begin{equation}} \def\eeq{\end{equation}} \def\hph{\hphantom}
\def\be{\begin{equation}} \def\ee{\end{equation}} \def\r#1{$^{[#1]}$}
\def\n{\noindent} \def\ni{\noindent} \def\pa{\parindent} 
\def\hs{\hskip} \def\vs{\vskip} \def\hf{\hfill} \def\ej{\vfill\eject} 
\def\cl{\centerline} \def\ob{\obeylines}  \def\ls{\leftskip}
\def\underbar#1{$\setbox0=\hbox{#1} \dp0=1.5pt \mathsurround=0pt
   \underline{\box0}$}   \def\ub{\underbar}    \def\ul{\underline} 
\def\f{\left} \def\g{\right} \def\e{{\rm e}} \def\o{\over} 
\def\vf{\varphi} \def\pl{\partial} \def\cov{{\rm cov}} \def\ch{{\rm ch}}
\def\la{\langle} \def\ra{\rangle} \def\EE{e$^+$e$^-$}
\def\bitz{\begin{itemize}} \def\eitz{\end{itemize}}
\def\btbl{\begin{tabular}} \def\etbl{\end{tabular}}
\def\btbb{\begin{tabbing}} \def\etbb{\end{tabbing}}
\def\beqar{\begin{eqnarray}} \def\eeqar{\end{eqnarray}}
\def\\{\hfill\break} \def\dit{\item{-}} \def\i{\item} 
\def\bbb{\begin{thebibliography}{9} \itemsep=-1mm}
\def\ebb{\end{thebibliography}} \def\bb{\bibitem}
\def\bpic{\begin{picture}(260,240)} \def\epic{\end{picture}}
\def\akgt{\noindent{\bf Acknowledgements}}
\def\fgn{\noindent{\bf\large\bf Figure captions}}
\bd
\pge
\bcc
\vskip-2.5cm
\hskip11cm{\large HZPP-9909}

\hskip11cm{\large July 30, 1999}

\vskip1.5cm

{\Large The Influence of Multiplicity Distribution }

{\Large on the Erraticity Behavior of Multiparticle Production
\footnote{ \ This work is supported in part by the 
Natural Science Foundation of China \\ 
\null{} \hs0.6cm (NSFC) under Grant No.19575021.}}
\vskip0.5cm

{Liu Zhixu \ \ \ \ \ \ \ Fu Jinghua \ \ \ \ \ \ \ Liu Lianshou }

{\small  Institute of Particle Physics, Huazhong Normal University, 
Wuhan 430079 China}

\vskip0.0cm

{\small Tel: 027 87673313 \qquad FAX: 027 87662646 
\qquad email: liuls@iopp.ccnu.edu.cn}
\date{ }
\begin{minipage}{125mm}
\vskip 1.5cm
\begin{center}{\Large Abstract}\end{center}
\vskip 0.0cm
\ \ \ \  The origin of the erraticity behaviour observed recently in the 
experiment is studied in some detail. The negative-binomial distribution
is used to fit the experimental multiplicity distribution. It is shown that, 
with the multiplicity distribution taken into account, the experimentally 
observed erraticity behaviour can be well reproduced using a flat
probability distribution. The dependence of erraticity behaviour on the 
width of multiplicity distribution is studied.

\end{minipage}
\end{center}
\vskip 1in
{\large PACS number: 13.85 Hd
\vskip0.2cm

\ni
Keywords: Multiparticle production, \ Negative-binomial distribution

\hs1.8cm Erraticity}

\npg \pgn \spg{2}

Since the finding of unexpectedly large local fluctuations 
in a high multiplicity event recorded by the JACEE 
collaboration~\cite{JACEE}, the investigation of non-linear phenomena in 
high energy collisions has attracted much attention~\cite{Kittel}. The
anomalous scaling of factorial moments, defined as 
\beqar 
F_q &=& 
    \frac{1}{M}\sum\limits_{m=1}^{M}
    \frac{\la n_m(n_m-1) \cdots (n_m-q+1)\ra} 
    {\la n_m \ra ^q  }   
\eeqar
at diminishing phase space scale or increasing division number 
$M$ of phase space~\cite{BP}:
\beqar 
F_q \propto M^{-\phi_q},
\eeqar
called intermittency (or fractal) has been proposed for this purpose.
The average $\la \cdots \ra$ in Eq.(1) is over 
the whole event sample and $n_m$ is the number of particle falling in the
$m$th bin.  This kind of anomalous scaling has been observed successfully
in various experiments~\cite{NA22}\cite{NA27}.

A recent new development along this direction is the event-by-event
analysis~ \cite{BiaZiaja}\cite{Liu}. An important step in this kind of 
analysis was made by  Cao and  Hwa~\cite{CaoHwa}, who first pointed out the 
importance of the fluctuation in event space of the event factorial moments
defined as
\beqar  
F_q^{({\rm e})} &=& \frac
    { \frac{1}{M}\sum\limits_{m=1}^{M} n_m(n_m-1) \cdots (n_m-q+1)} 
    { \f( \frac{1}{M}\sum\limits_{m=1}^{M} n_m\g)^q} .
\eeqar
Its fluctuations from event to event can be quantified
by its normalized moments as:
\begin{equation} 
C_{p,q}=\la \Phi_q^p\ra, \quad \Phi_q= {F_q^{(e)}} \f/ 
\la F_q^{(e)}\ra \g.,
\end{equation}
\noindent and by $dC_{p,q}/dp$ at $p=1$:
\beqar   
\Sigma_q=\la \Phi_q \ln \Phi_q \ra .
\eeqar
If there is a power law behavior of the fluctuation as 
division number goes to infinity, or as resolution $\delta=\Delta/M$ 
goes to very small, {\rm i.e.},
\begin{equation}   
C_{p,q}(M) \propto M^{\psi_q(p)},
\end{equation}
then the phenomenon is referred to as erraticity~\cite{hwa}.
The derivative of exponent $\psi_q(p)$ at $p=1$ 
\begin{equation}   
\mu_q=\f.\frac{d}{dp}\psi_q(p)\g|_{p=1} = 
             \frac{\partial \Sigma_q}{\partial \ln M}.
\end{equation}
describes the anomalous scaling property of fluctuation-width and is called 
entropy index. 

The erraticity behaviour of multiparticle final states as described above 
has been observed in the experimental data of 400 GeV/$c$ pp collisions 
from NA27~\cite{WSS}.  However, it 
has been shown~\cite{ZGKX} that the single event factorial moment as defined
in Eq.(3), using only the horizontal average over bins, cannot eliminate
the statistical fluctuations well, especially when the multiplicity is low.
A preliminary study shows that
the experimentally observed phenomenon~\cite{WSS} can be reproduced by using
a flat probability distribution with only statistical fluctuations~\cite{ZGKX}.
This result is preliminary in the sense that it has fixed the multiplicity to
9 while the multiplicity is fluctuating in the experiment and has an average of
$\la n_{\rm ch}\ra = 9.84$~\cite{NA27avn}. Since the erraticity phenomenon
is a kind of fluctuation in event space and depends strongly on the 
multiplicity, the fluctuation in event space of the multiplicity is expected
to have important influence on this phenomenon. 

In this letter this problem is discussed in some detail. The
negative binomial distribution~\cite{NBD} 
will be used to fit the experimental multiplicity 
distribution~\cite{NA27avn}.  Putting the resulting 
multiplicity distribution into a flat-probability-distribution 
model, the erraticity behaviour is obtained and compared with the 
experimental data. The consistency of these two shows 
that the erraticity behaviour observed in the 400 GeV/$c$ pp collision data
from NA27 is mainly due to statistical fluctuations.

The negative-binomial distribution is defined as~\cite{NBD}
\beq 
P_n=\f(\matrix{n+k-1 \cr n} \g)
\f(\frac{\bar n/k}{1+\bar n/k}\g)^n \frac{1}{(1+\bar n/k)^k} ,
\eeq
where $n$ is the multiplicity, $\bar n$ is its average over event sample,
$k$ is a parameter related to the second order scaled moment
$C_2 \equiv \la n^2\ra/\la n^2\ra$ through~\cite{NBD}
\beq 
C_2-1= \frac{1}{\bar n} + \frac{1}{k} .
\eeq

Using Eq.(8) to fit the multiplicity distribution of 400 GeV/$c$ pp
collision data from NA27, we get the parameter $k=12.76$. The result of
fitting is shown in Fig.1. It can be seen that the fitting is good.

Then we take a flat (pseudo)rapidity distribution, i.e. let the probability 
for a particle to fall into each bin be equal to $p_m=1/M$ when the  
(pseudo)rapidity space is divided into $M$ bins. This means that there 
isn't any dynamical fluctuation.

Let the number $N$ of particles in an event be a random number distributed
according to the negative binomial distribution Eq.(8) with 
$\bar n =9.84, k=12.76$.  Put these $N$ particles into the $M$ bins according 
to the Bernouli distribution

\npg
\beq  
 B(n_1,n_2,\cdots ,n_M |p_1,p_2,\cdots ,p_M)=
        \frac {N!}{n_1! \cdots n_M!} p_1^{n_1}\cdots p_M^{n_M} ,
\eeq
$$  \sum_{m=1}^M n_m =N . $$
In total 60000 events are simulated in this way and the resulting $C_{p,q}$
are shown in Fig.2 together with the experimental data of 400 GeV/$c$ pp 
collisions from NA27. It can be seen from 
the figures that the model results are consistent with the data, showing 
that the erraticity phenomenon observed
in this experiment is mainly due to statistical fluctuations.

In order to study the relation of erraticity behaviour with the width of 
multiplicity
distribution, the same calculation has been done for the cases:
$\bar n =9, k=0.1, 0.5, 1.0, 2.25, 4.5, 9, 18$.  These values of $k$ 
corresponds to diminishing width of distribution with 
$C_2= 11.1, 3.11, 2.11,$
$1.56, 1.33, 1.22, 1.17$ respectively, cf. Fig.3.  
The resulting ln$C_{p,2}$ and
$\Sigma_2 $ as function of ln$M$ are shown in Fig.4 and Fig.5.

It can be seen from the figures that the moments
$C_{p,2}$ for different $p$ separate farther and the 
characteritic function $\Sigma_2$
becomes larger when the value of $k$ is smaller. This means that the 
single event factorial moments fluctuate stronger in event space
when the width of multiplicity distribution is wider. On the other hand,
the straight lines obtained from fitting the last three points of $\Sigma_2$
versus ln$M$ are almost parallel for different $k$, and their
slopes --- the entropy indices $\mu_2$, 
which is the characteristic quantity of erraticity 
are insensitive to the width of multiplicity distribution. 

In summary, the multiplicity distribution of 400 GeV/$c$ pp collision data
from NA27 has been fitted to the negative binomial distribution.
Taking this multiplicity distribution into account, the erraticity 
phenomenon in a model without any dynamical fluctuation, i.e. with a flat 
probability distribution, has been studied. The resulting moments $C_{p,q}$
turn out to fit the experimental data very well. This shows that the
erraticity phenomenon observed in this experiment is mainly due to 
statistical fluctuations. 

The dependence of erraticity phenomenon on the width of multiplicity 
distribution is examimed. It is found that  the
fluctuation of single event factorial moments in event space becomes 
stronger --- $C_{p,2}$ and $\Sigma_2$ become larger ---
when the width of multiplicity distribution is wider. On the other 
hand, the entropy index $\mu_2$ depends mainly on the average multiplicity
and is insensitive to the width of multiplicity distribution. 

\newpage
\def\Journal#1#2#3#4{{#1} {\bf #2} (#3) #4}
\def\NCA{\em Nuovo Cimento} \def\NIM{\em Nucl. Instrum. Methods}
\def\NIMA{{\em Nucl. Instrum. Methods} A} \def\NPB{{\em Nucl. Phys.} B}
\def\PLB{{\em Phys. Lett.}  B} \def\PRL{\em Phys. Rev. Lett.}
\def\PRD{{\em Phys. Rev.} D} \def\ZPC{{\em Z. Phys.} C}
\def\PRE{{\em Phys. Rev.} E} \def\PRC{{\em Phys. Rev.} C} 

\newpage

\ni{\Large\bf Figure Captions}
\vs1cm

{\pa=0pt{\ls=15mm\rightskip15mm
\hs-15mm
{\bf Fig.1} \ Fitting of the multiplicity distribution of 400 GeV/$c$ pp
collision data to negative binomial distribution. Data taken from Ref.[12].
\par}}

\vs1cm
{\pa=0pt{\ls=15mm\rightskip15mm
\hs-15mm
{\bf Fig.2} \ The 
moments $C_{p.2}$ from a flat probability distribution model with the
multiplicity distribution taken into account, as compared with the 400 GeV/$c$ 
pp collision data taken from Ref.[12].
\par}}

\vs1cm
{\pa=0pt{\ls=15mm\rightskip15mm
\hs-15mm
{\bf Fig.3} \ The negative binomial distribution with different values of
parameter $k$. The average multiplicity is $\bar n = 9$.
\par}}

\vs1cm
{\pa=0pt{\ls=15mm\rightskip15mm
\hs-15mm
{\bf Fig.4} \ The dependence of ln$C_{p,2}$ on ln$M$ in the flat 
probability distribution model,
taken the negative-binomial type multiplicity distribution into account.  
The parameter $k$ takes different vaues as shown in the figure. 
The average multiplicity is $\bar n = 9$.
\par}}

\vs1cm
{\pa=0pt{\ls=15mm\rightskip15mm
\hs-15mm
{\bf Fig.5} \ The dependence of $\Sigma_2$ on ln$M$ in the flat probability 
distribution model,
taken the negative-binomial type multiplicity distribution into account.  
The parameter $k$ takes different vaues as shown in the figure. 
The average multiplicity is $\bar n = 9$.
\par}}

\newpage

\baselineskip 0.18in

\begin{picture} (260,240) 
\put(75,-70)   
{\epsfig{file=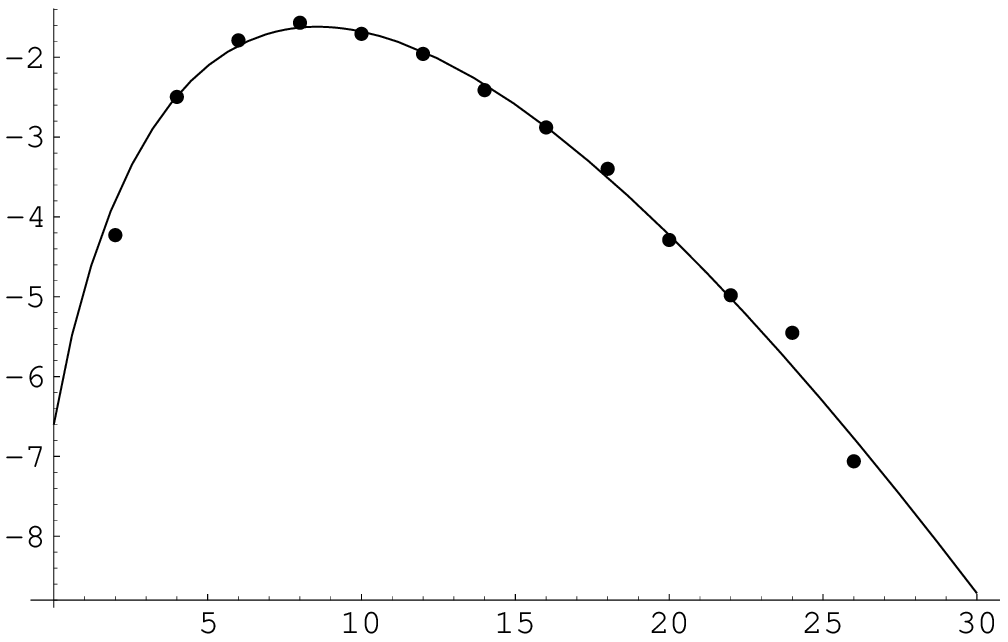,bbllx=0cm,bblly=0cm,
           bburx=8cm,bbury=6cm}}  
\end{picture}
\vskip-6cm\hskip1cm ln$P(n_{\rm ch})$

\vskip6.2cm \hskip12cm $n_{\rm ch}$

\vskip2.0cm
\cl{Fig. 1}

\begin{picture} (260,240) 
\put(-75,-310)   
{\epsfig{file=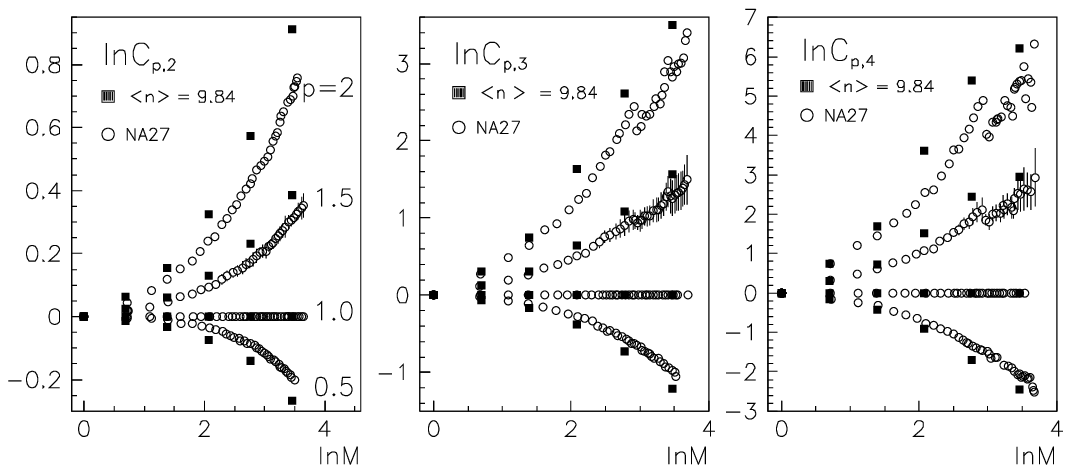,bbllx=0cm,bblly=0cm,
           bburx=8cm,bbury=6cm}}  
\end{picture}
\vs0.5cm
\cl{Fig. 2}

\newpage

\begin{picture} (260,240) 
\put(0,-220)   
{\epsfig{file=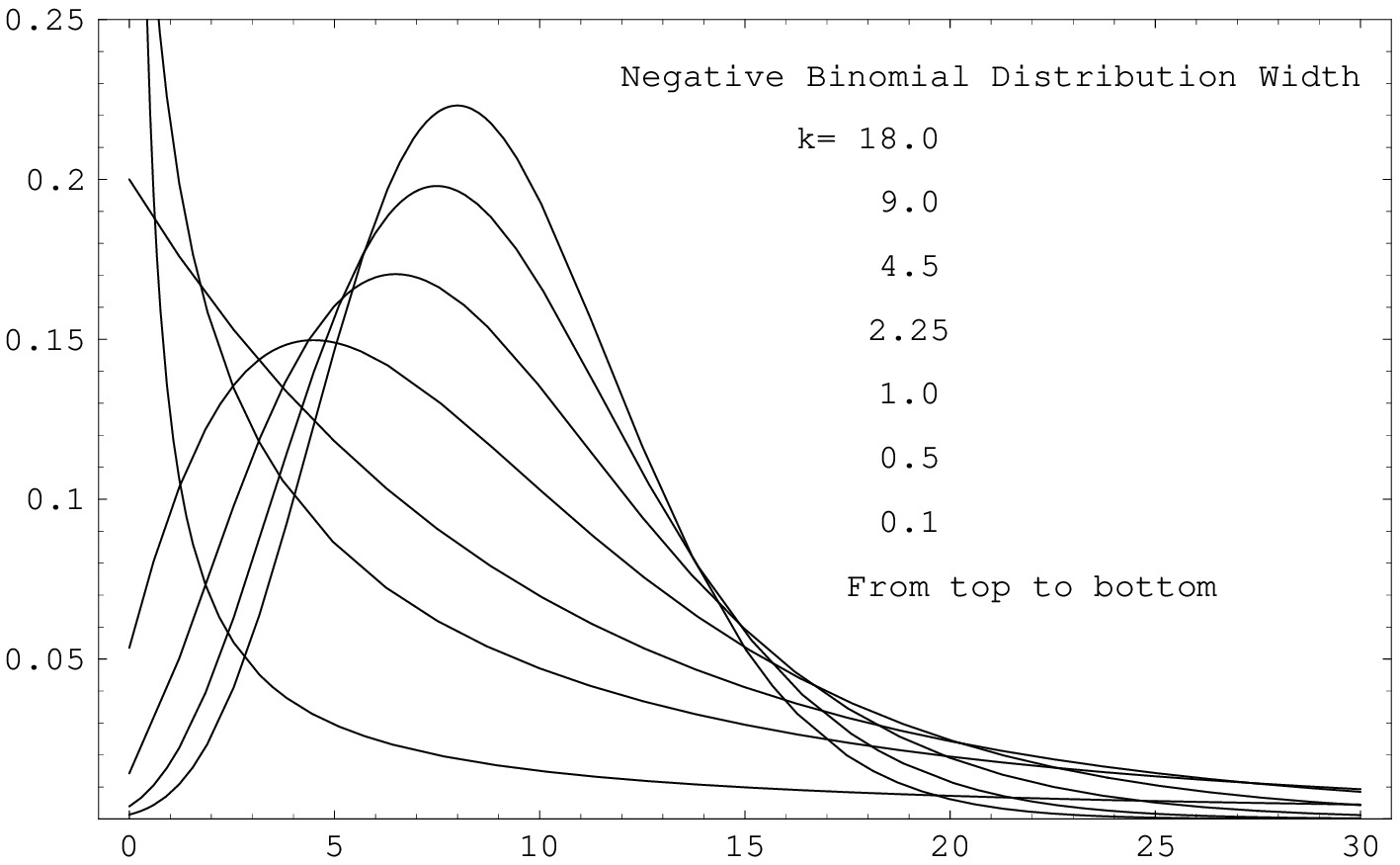,bbllx=0cm,bblly=0cm,
           bburx=8cm,bbury=6cm}}  
\end{picture}
\vskip-3cm ln$P(n_{\rm ch})$

\vskip9.4cm \hskip14cm $n_{\rm ch}$

\vs2cm
\cl{Fig. 3}

\newpage
\begin{picture} (260,240) 
\put(-75,-420)   
{\epsfig{file=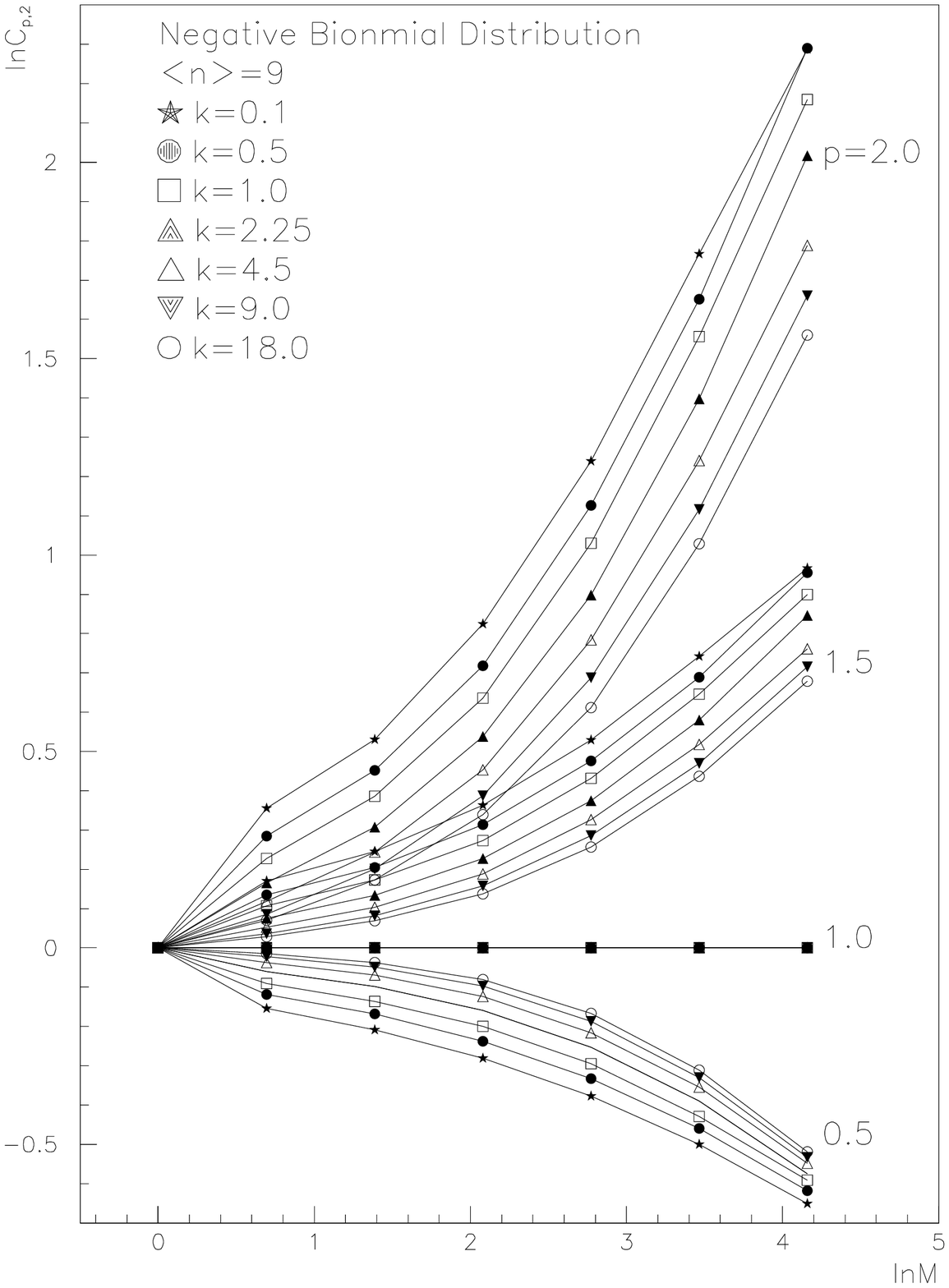,bbllx=0cm,bblly=0cm,
           bburx=8cm,bbury=6cm}}  
\end{picture}

\vs13cm
\cl{Fig. 4}

\newpage
\begin{picture} (260,240) 
\put(-75,-420)   
{\epsfig{file=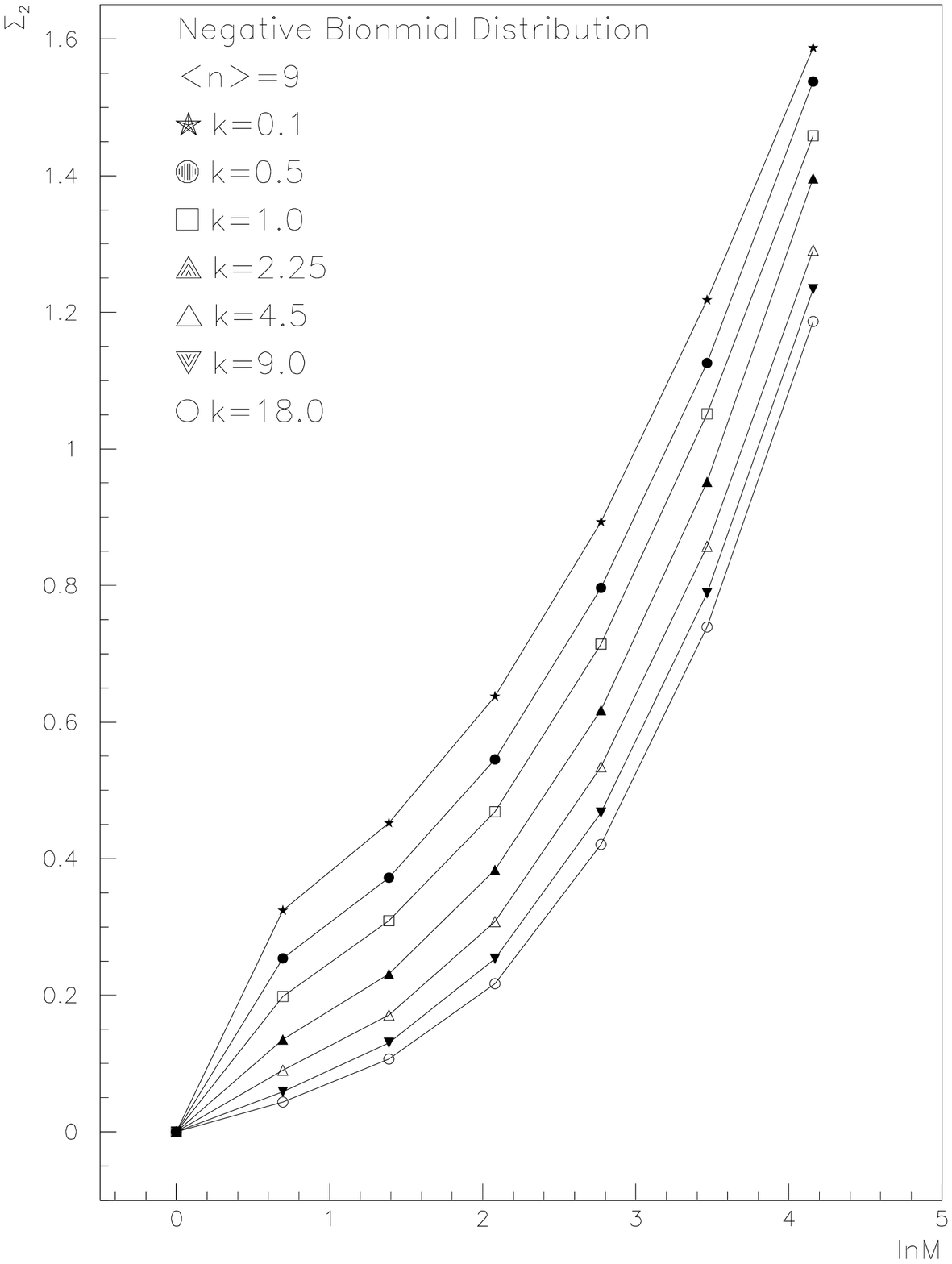,bbllx=0cm,bblly=0cm,
           bburx=8cm,bbury=6cm}}  
\end{picture}

\vs13cm
\cl{Fig. 5}

\ed